\documentclass[%
 reprint,
amsmath,amssymb,
aps, 
]{revtex4-2}

\usepackage{graphicx}
\usepackage{dcolumn}
\usepackage{bm}
\usepackage{comment}

\begin{document}

\preprint{APS/123-QED}
\title{Unraveling the role of excitons in the near ideal performance of perovskite light emitting diodes}

\author{Pradeep R. Nair}
 \email{prnair@ee.iitb.ac.in}
\affiliation{ Department of Electrical Engineering,
 Indian Institute of Technology Bombay, Mumbai, India\\
}%




\date{\today}

\begin{abstract}
Recent reports indicate that perovskite based light emitting diodes (LEDs) have achieved an external quantum efficiency (EQE) of 32\% - rather an internal quantum efficiency close to 100\%. Much of this improved performance is attributed to the role of excitons. While the experimental trends are encouraging, the  recombination parameters estimated through extensive curve-fitting of photoluminescence (PL) transients are not amenable to reasonable interpretations. In view of the same, through a detailed analysis of free carrier - exciton dynamics in perovskite optoelectronic materials, here we identify a coherent scheme to unambiguously back extract the relevant parameters. The model predictions compare well with the recent experimental results on perovskite LEDs with record EQE thus quantifying the role of excitons. Importantly, this work identifies a physics aware scheme for the design of experiments tailored towards consistent exploration of underlying physical mechanisms and hence could enable a synergistic optimization of process technology and device performance.   
\end{abstract}

\maketitle


\section{\label{sec:intro}Introduction}
Peorvskite optoelectronic devices like solar cells and LEDs continue their unmatched progress towards commercialization with impressive improvement in performance \cite{Yoo2021,Shen2024,Sun2023,Kim2022,longi,Li2024}. Importantly, as per recent reports \cite{Li2024}, external quantum efficiency of perovskite based LEDs now approach 32\% - which indicates that nearly 100\% of the contact injected carriers recombine through radiative means. This is a significant achievement and the increased radiative recombination is attributed to excitons. In fact, several reports are available in recent literature on the influence of excitons in perovskite optoelectronics \cite{Simbula,Zhao2023,Mariano2020}.\\

Detailed knowledge of the recombination mechanisms and the associated parameters are central to the understanding and optimization of all optoelectronic devices \cite{stranks_review_recombination,park_review,motti_review,Fakharuddin2022,sumanshu_apl,nair_acsEL}. At a material or thin film level, several techniques including time resolved PL measurements (TRPL) are widely used to back extract the recombination parameters \cite{OPTP_TRMC_review,OPTP_review_chargedynamics,TAS_review,jacs_transient,diffuse_reflectance,TRMC_ACSPhtonics,Nair2023}. Our recent work achieved the same at a  device level through terminal current-voltage characteristics \cite{Hossain2024}. 
However, there exist several gaps in the back extraction strategy and interpretation of parameters using transient PL measurements.  To list a few: (i) what features in TRPL could be unambiguously and directly attributed to excitons? (ii) Is the effective monomolecular recombination rate also influenced by excitons (as claimed in literature \cite{herz_exciton,Li2024})?. (iii) How informative are the parameters estimated via rigorous curve-fitting of TRPL data? For example, recent reports indicate \cite{Li2024} that the bimolecular recombination rate could be of the order of $10^{-9} \,cm^3/s$ in perovskite LEDs. For a comparable band gap (i.e., $E_g \approx 1.6 \,eV)$, the fundamental radiative recombination rate, which is consistent with the Shockley-Queisser analysis \cite{SQ}, is much smaller than $10^{-11} \,cm^3/s$ (see ref. \cite{bolink_eqeel,sumanshu_apl}). Can excitons account for the several orders of magnitude improvement in effective $k_2$? (iv) Is it possible for such back extracted parameters (i.e., from TRPL) to anticipate the steady state PL Quantum Efficiency (PLQE) measurements?\\

Many of the above conflicts/puzzles arise due to the lack of appropriate asymptotic analytical solutions that allows lucid interpretation of experiments. Absence of the same necessitates the need for complex data extraction protocols whose results often do not convey insights in a simple manner. In this manuscript, we provide a coherent and consistent theoretical analysis of free carrier - exciton dynamics with analytical solutions that allow facile back extraction of relevant parameters from appropriate regimes. This physics aware methodology addresses all puzzles listed above and allows further design of experiments as described in the following sections. Curiously, in contrast to a phrase attributed to Aristotle, here we find that the sum of the parts (i.e., analysis using predictive models over multiple regimes) is indeed more informative than the whole (i.e., brute force curve-fitting).
\section{Curve-fitting PL transients}
As mentioned before, carrier lifetimes play a significant role in the performance of optoelectronic devices. Simple schemes to estimate the carrier lifetimes at a material or thin film level involves curve-fitting PL transients with single exponentials, bi-exponentials, stretched exponentials or a combination thereof \cite{kamat_bi,Lu_stretchexp,amrita_PL,gingerkinetics}. Other schemes involve analysis of PL transients using the following rate equation (also known as ABC model \cite{herz_exciton,deQuilettes2019})
\begin{equation}
     \frac{\partial n}{\partial t} = G - k_1n-k_2n^2-k_3n^3 \\
    \label{eq:sim_rate}
\end{equation}
where $n$ is the free carrier density and $G$ is the carrier generation rate. The parameters $k_1$, $k_2$, and $k_3$ denote the monomolecular, bimolecular, and Auger recombination processes. These recombination parameters are widely obtained through curve-fitting of PL transients with eq. \ref{eq:sim_rate} (i.e., during pulse OFF period with $G=0$). In addition, the PLQE is given as
\begin{equation}
     PLQE = \frac{k_2n^2}{k_1n+k_2n^2+k_3n^3}  \\
    \label{eq:sim_PLQE}
\end{equation}
where $n$ is obtained as solution of eq. \ref{eq:sim_rate} under steady state conditions (with $G  \neq 0$).

\begin {figure} [ht!]
  \centering
    \includegraphics[width=0.45\textwidth]{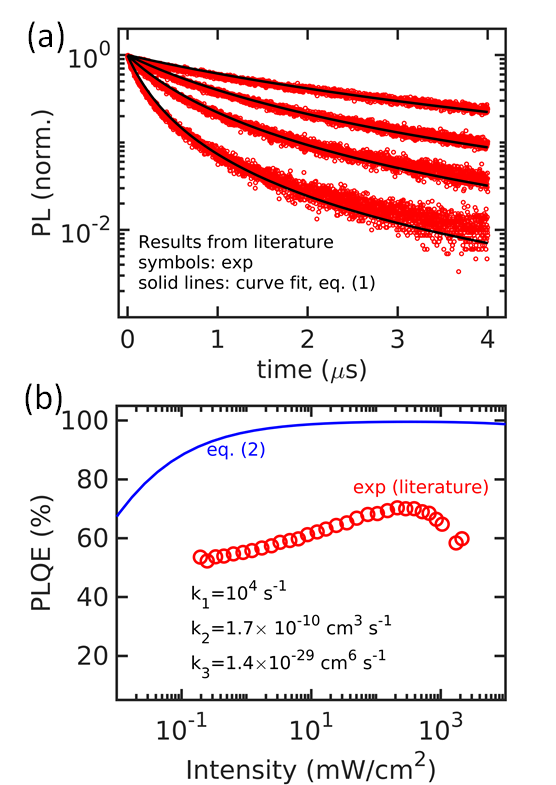}
    \caption{\textit{Relevance of parameter extraction through curve-fitting of PL transients. (a) Experimental TRPL (symbols) and curve-fitting results (solid lines) as reported for the 'control' sample in ref. \cite{Li2024}. (b) PLQE of the same sample (symbols). The solid line indicates the PLQE estimated using eq. \ref{eq:sim_PLQE} with the parameters listed, as reported in ref. \cite{Li2024}. 
     }}
\label{exp_tran1}
\end{figure}

A natural question arises at this stage: How good are the parameters extracted from TRPL measurements (i.e., through curve-fitting of eq. \ref{eq:sim_rate})? Rather, can they anticipate the PLQE trends? Figure \ref{exp_tran1} shows experimental results from recent literature \cite{Li2024}. Part (a) shows the PL transients while part (b) shows the PLQE. This perovskite sample reported a maximum PLQE of around 70\% and the fully finished LED reported a maximum EQE of around 20\%. The solid lines in Fig. \ref{exp_tran1}a indicate curve-fit using eq. (1), as reported in the same publication \cite{Li2024} with $R^2>0.99$, which apparently indicates a good fit. The back extracted parameters, as reported in ref. \cite{Li2024}, are mentioned in Fig. \ref{exp_tran1}b. \\

The solid line in Fig. \ref{exp_tran1}(b) indicates our attempt to  predict the PLQE using the back extracted parameters from TRPL (i.e., using eq. \ref{eq:sim_PLQE}). Surprisingly, even though the fit to TRPL looks excellent (with $R^2>0.99$, as reported \cite{Li2024}), the back extracted parameters are found to be inadequate to anticipate the PLQE trends - both in terms of the magnitude as well as in terms of its dependence on intensity. The experimental data shows a near logarithmic dependence on intensity while the predicted trends are rather independent of intensity (over the same range). Li et al. \cite{Li2024} also reports results for a sample with PLQE of 96\%. Their analysis of the corresponding TRPL indicates that the $k_2$ increases by about two orders of magnitude, while the $k_1$ also increases by an order of magnitude (as compared to the sample with PLQE of 70\%). This is a very puzzling trend - are excitons expected to increase both $k_1$ and $k_2$?  The mismatch  between theory and experiments is more worrisome in view of prevalent impression that $k_1$ could be due to both monomolecular free carrier recombination as well as exciton recombination \cite{herz_exciton, Li2024}. Given such an interpretation, is it difficult to umabiguously  quantify the excitonic contributions to the PLQE of perovskite LEDs. 
\section{Proposed Methodology}
There are several shortcomings to curve-fitting of PL transients using eq. \ref{eq:sim_rate}. For example, eq. \ref{eq:sim_rate} do not explicitly account for phenomena like carrier trapping and exciton dynamics. A detailed model to address the influence of these diverse phenomena is provided in Methods section (see eq. \ref{eq:traprate}). The general scheme, under certain assumptions like  (a) quasi-steady state between excitons and free carriers, (b) quasi-steady state conditions for traps, (c) mid-level traps, and (d) $n=p$ (where $n$ and $p$ denotes the free electron and free hole densities, respectively) reduces to the following simplified formalism for carrier - exciton dynamics and PL as
\begin{subequations}
\begin{align}
    (1+\frac{2n}{n_{ep}}) \frac{\partial n}{\partial t} &= G - k_1n - k_2^{'}n^2-k_3n^3\\
    k_2^{'}&=k_2+\frac{k_x}{n_{ep}}\\
    PL&=(k_{2,rad}+k_x/n_{ep})n^2
    \end{align}
    \label{eq:eff_rate1}
\end{subequations}

The bimolecular recombination rate ($k_2^{'}$) consist of radiative and non-radiative components  ($k_2=k_{2,rad}+k_{2,nonrad}$) and excitonic contribution ($k_x/n_{ep}$). Here $k_x$ denotes the radiative decay rate of excitons (i.e., with photon emission). Under quasi-steady conditions between free carriers and excitons, we have  $n^2=n_x\times n_{ep}$ - the Saha equilibrium (where $n_x$ is the density of excitons, and $n_{ep}$ depends on the exciton binding energy, see eq. \ref{eq:saha}, Methods section). It is instructive to note that exciton influences the effective bimolecular recombination rate ($k_2^{'}$) and not the mono-molecular recombination rate ($k_1$). As such, there are several unknown parameters to be estimated ($k_1$, $k_2$, $k_3$, and $k_x$). Based on eq. \ref{eq:eff_rate1}, below we identify a consistent back extraction strategy for the relevant parameters.\\

\textbf{A. Back extraction of $k_1$}:
During pulse OFF transients (i.e., with $G=0$),  for small $n$ and dominant monomolecular recombination, eq. \ref{eq:eff_rate1}a reduces to $\partial n/\partial t \approx -k_1n$. Carrier transients are then given as $n \propto e^{-k_1t}$. Accordingly, the time constant associated with PL transients is given as 
\begin{equation}
 \tau^{-1}=2k_1  
 \label{k1}
\end{equation}
Hence the parameter $k_1$ could be directly back extracted from the time constants of PL transients.\\

\textbf{B. Back extraction of $k_2$}:
For OFF transients with moderate $n$ ($n<n_{ep}$), with negligible Auger recombination, eq. \ref{eq:eff_rate1}a reduces to $\partial n/\partial t \approx -k_1n -k_2^{'}n^2$. It can be shown that the time constant associated with the early phase of  PL transients is given as
  \begin{equation}
 \tau^{-1}=2(k_1+k_2^{'} \times n(0))
   \label{k2}
 \end{equation}
 Equation \ref{k2} indicates that both the parameters $k_1$ and $k_2^{'}$ can be estimated through intensity dependent transient measurements - i.e., by measuring the early phase time constants for several excitation intensities and plotting the same against the initial carrier density $n(0)$. From such a plot, $k_1$ can be obtained as the intercept and $k_2^{'}$ as the slope. The parameter $k_2^{'}$, given by eq. \ref{eq:eff_rate1}, is also influenced by excitonic contribution.\\
 
\textbf{C. Back extraction of $k_x$}:
For $n$ comparable to $n_{ep}$ and dominated by bimolecular recombination, eq. \ref{eq:eff_rate1}a reduces to $2\partial n/\partial t \approx -k_2^{'}n_{ep}n$. Under such conditions,  the time constant associated with PL transients is given as 
\begin{equation}
 \tau^{-1} = k_x + k_2 \times n_{ep} 
 \label{kx}
\end{equation}
Accordingly, the parameter $k_x$ and $k_2$ can be directly estimated from the time constants of different regimes of PL transients (as indicated by eqs. \ref{k2}-\ref{kx}).\\

\textbf{D. Back extraction of $k_3$}:
The contribution of Auger recombination becomes appreciable only in the presence of large carrier densities. For large $n$ and negligible monomolecular recombination, eq. \ref{eq:eff_rate1}a reduces to $2\partial n/\partial t \approx -k_2^{'}n_{ep}n-k_3n_{ep}n^2$. The corresponding time constant of the early phase of transients is given as 
\begin{equation}
 \tau^{-1} = k_2^{'}n_{ep} + k_3n_{ep} \times n(0) 
 \label{k3}
\end{equation}
which indicates that both $k_2^{'}$ and $k_3$ can be obtained from the transients under large excitation intensities (i.e., from the intercept and slope of $\tau^{-1}$ vs. $n(0)$).\\

Equations \ref{k1}-\ref{k3} identify several regimes and corresponding $\tau$ in terms of the parameters associated with free carrier - exciton dynamics. Among them, eqs. \ref{k1} and \ref{kx} indicate that the corresponding rates are independent of the excitation level while eqs. \ref{k2} and \ref{k3} predict a linear dependence on excitation levels (i.e., $n(0)$). Hence, it is indeed possible to design intensity dependent experiments to uniquely back extract all the relevant parameters in an unambiguous manner - without resorting to curve-fitting the entire transients.\\

\textbf{E. Insights from PLQE}:
In addition to the transients, the variation of steady state PL with illumination intensity is of significant interest - especially from a device perspective. Several trends are readily available in this case as well. For example,  eq. \ref{eq:eff_rate1}a indicates that $n \propto G$ when monomolecular recombination dominates under steady state conditions. Consequently, $PL \propto G^2$ under such conditions and the $PLQE \propto G$. Similarly, we find that $PL \propto G $ when $k_{2}^{'}$ dominates - i.e.,  the steady state PL signal will always be proportional to $G$ - regardless of whether dominated by free carrier radiative recombination or dominated by excitonic effects. Under such conditions the PLQE will be rather independent of G. Further, under dominant Auger recombination, we find $PL \propto G^{2/3}$ and the PLQE will vary as $G^{-1/3}$.\\

In addition to the scaling trends with $G$, the functional dependence of maximum PLQE (denoted as $PLQE_{max}$) on various parameters can be directly obtained. Under steady state photo excitation, eq. \ref{eq:eff_rate1}a indicates that  $G=k_1n+k_{2}^{'}n^2+k_3n^3$. Further, using eq. \ref{eq:PL_max1}a, we find that $PLQE_{max}$ is attained under the conditions $n_{PL}=(k_1/k_3)^{1/2}$ - an interesting result as this critical carrier density is not influenced by either radiative recombination or excitons. Accordingly, we have
 \begin{subequations}
\begin{align}    
    PLQE &= \frac{(k_{2,rad}+k_x/n_{ep})n^2}{k_1n+k_{2}^{'}n^2+k_3n^3}\\
    PLQE_{max} & = \frac{k_{2,rad}+k_x/n_{ep}}{k_{2,rad}+k_{2,nonrad}+k_x/n_{ep}+2\sqrt{k_1k_3}}
\end{align}    
    \label{eq:PL_max1}
\end{subequations}

Apart from its relevance to explore the parameters listed in eq. \ref{eq:eff_rate1} and \ref{eq:PL_max1}, PLQE can be used to back extract information regarding traps. In the presence of significant carrier trapping, the assumption $n=p$ is no longer valid. Such a regime can be explored through low level illumination intensities. As such, under low-level illumination with signifcant trapping, we have $n < p$ and $p \approx n_T$, where $p$ is the hole density and $n_T$ denotes the density of filled traps (see eq. \ref{eq:traprate}, Methods section). Dominant monomolecular recombination indicates that $G \approx k_1n$. By appropriately modifying eq. \ref{eq:PL_max1}a for $n\neq p$ (as indicated by eq. \ref{eq:traprate}), we find the PLQE at low  illumination intensities as 
\begin{equation}
PLQE_{low} \approx \frac{(k_{rad}+k_x/n_{ep})n_T}{k_1}             \label{eq:PLQY_l1}
\end{equation}
For single level traps, we have $n_T =  N_Tn/(n+n_1)$, where $N_T$ denotes the density of single level traps. For mid-level trapping centers, we have $n>n_1$ and hence the $PLQE_{low}$ will be independent of $G$ at low intensity levels. For shallow traps, we have $n<n_1$ and hence the $PLQE_{low}$ is expected to vary linearly with $G$. On the other hand, for uniform trap distribution, we can assume that all traps placed energetically lower than the quasi-Fermi level ($E_{Fn}$) are filled. With $E_{Fn}-E_i=kT/q \times log(n/n_i)$, we find $n_T=N_T (E_g/2+kT/q \times log (n/n_i))$, where $N_T$ denotes the uniform trap density, $n_i$ denotes the intrinsic carrier density, and $E_i$ denotes the intrinsic level. This result along with eq. \ref{eq:PLQY_l1} predicts that 
\begin{equation}
PLQE_{trap} = (k_{rad}+\frac{k_x}{n_{ep}})\frac{N_T}{k_1}(\frac{E_g}{2}+\frac{kT}{q}log (\frac{G}{k_1n_i}))                         
    \label{eq:PLQY_u1}
\end{equation}

The above equation indicates that $PLQE$ varies logarithmically with intensity for uniform trap distribution. Further, eq. \ref{eq:PLQY_u1} indicates that the slope of PLQE vs. log(Intensity) is given as $N_T(k_{rad}+k_x/n_{ep})kT/(qk_1)$ - which allows us to directly estimate $N_T$ from the PLQE results. Overall, we find: (a) mid-levels traps lead to PLQE being independent of illumination intensity, (b) shallow traps result in a linear dependence of PLQE on intensity, while (c) a logarithmic dependence of PLQE on intensity indicates the presence of uniform trap densities.\\ 

Equations \ref{k1}-\ref{eq:PLQY_u1} describe a comprehensive methodology which enables design of experiments and further analysis of PL results (both transients and steady state) to back extract various parameters like $k_1$, $k_x$, $k_{2,rad}$, $k_{2,nonrad}$, $k_3$, and trap density ($N_T$) - without resorting to curve-fitting PL transients with eq. \ref{eq:sim_rate}. As different regimes are identified with appropriate simple analytical models for $\tau$, this scheme could be very useful to analyze experiments.
\section{Application to experiments}
\begin {figure} [ht!]
  \centering
    \includegraphics[width=0.45\textwidth]{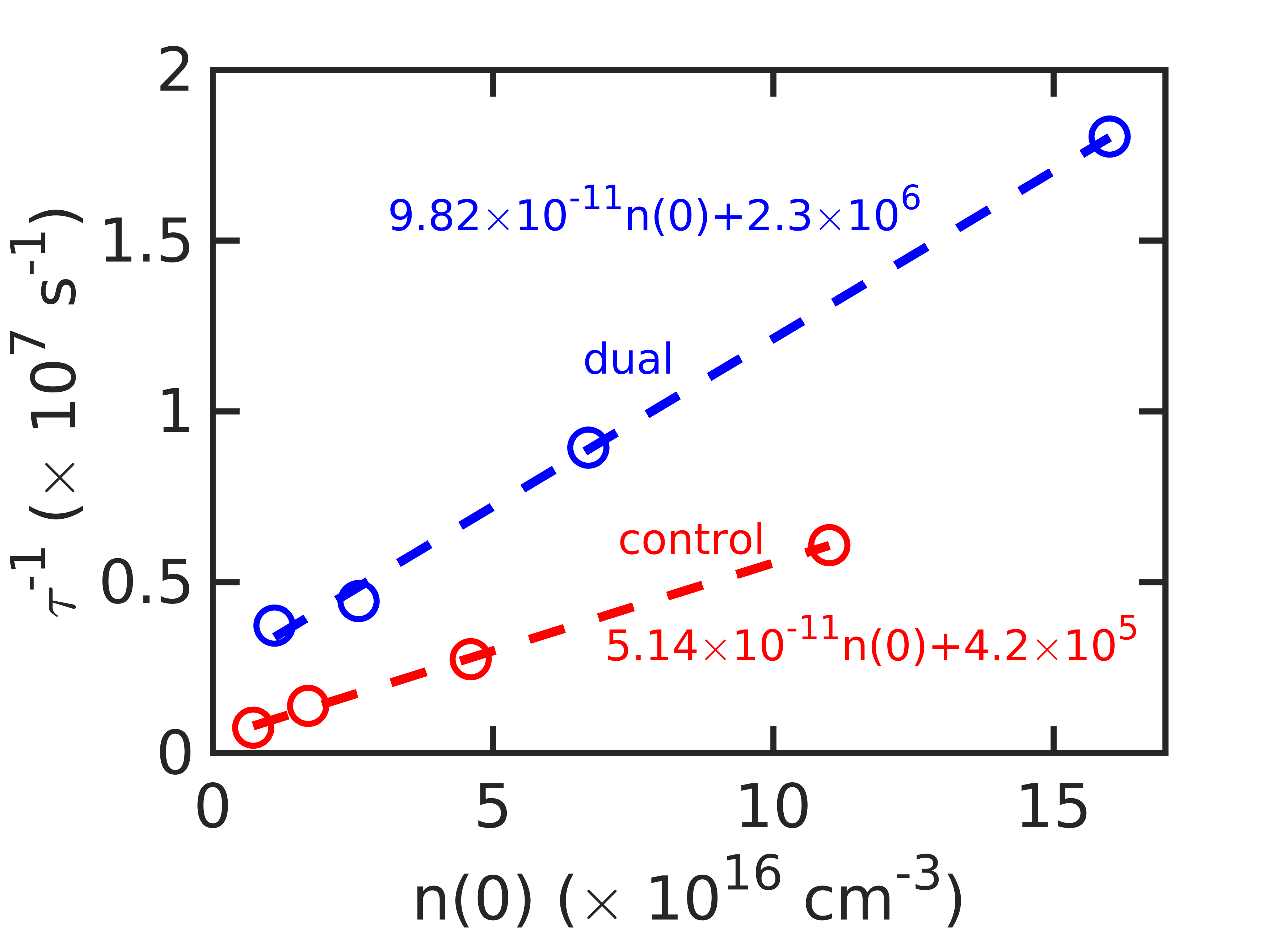}
    \caption{\textit{Analysis of time constants estimated from the PL transients reported in ref. \cite{Li2024}. The red and blue symbols represent control and dual samples, respectively. The expressions provided allow back extraction of parameters, as per eq. \ref{k1}-\ref{k3}.
     }}
\label{exp_fit}
\end{figure}

\begin {figure} [ht!]
  \centering
    \includegraphics[width=0.45\textwidth]{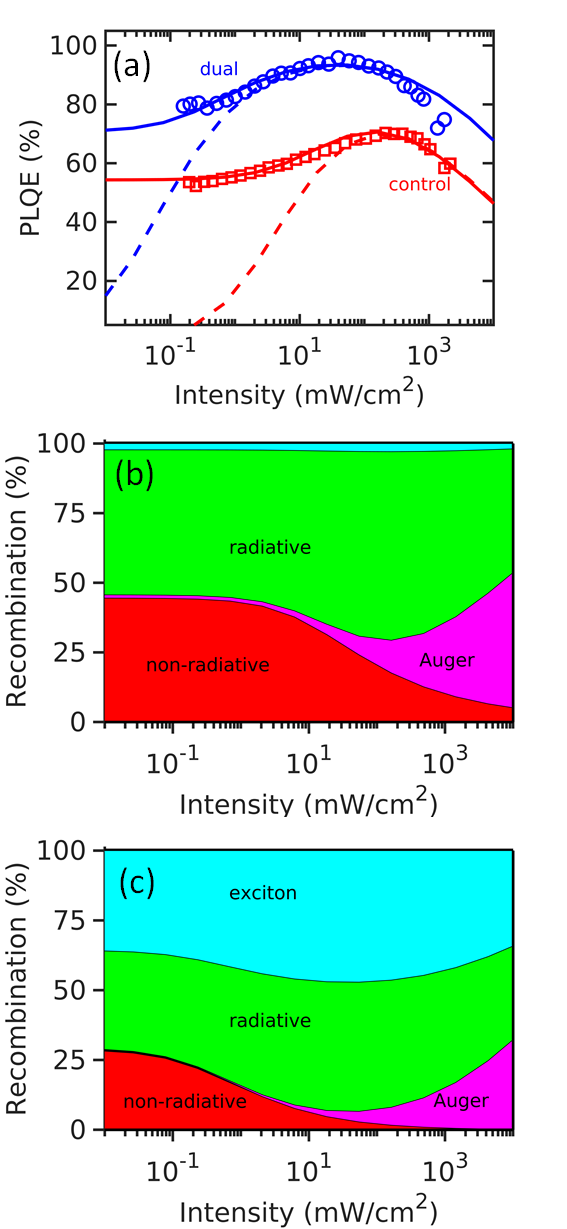}
    \caption{\textit{Comparison between proposed model and experimental results. (a) PLQE as a function of illumination intensity. The symbols correspond to experimental results from ref. \cite{Li2024}. The solid lines represent numerical solution of eq. \ref{eq:traprate} in the presence of carrier trapping. The dashed lines indicate theoretical predictions in the absence of carrier trapping. Parameters involved are provided in Table 1. Parts (b) and (c) shows the intensity dependent recombination components for the control sample and the dual sample, respectively. The excitonic component is significant and comparable to the free carrier radiative recombination for the dual sample with much reduced non-radiative and Auger recombination. These contribute to the near ideal performance for the dual sample.
     }}
\label{exp_PLQY}
\end{figure}

Here, we use our developed methodology to quantify the role of excitons on the PLQE of  perovskite LEDs with near ideal PLQE of 96\%. Two sets of data were reported by Li et al. with the same material \cite{Li2024}, however with different additives used in processing - (a) control sample with peak PLQE of 70\% (denoted as 'control' in this manuscript. Fig. 1 shows the TRPL and PLQE of this sample.) and (b) a sample processed with dual additives with PLQE of 96\%, denoted as 'dual' in this manuscript. Figure \ref{exp_fit} shows the time constants estimated from the experimental PL transients as a function of the generated carrier density. Specifically, the time constants were estimated using the data reported in Figs. 3a and 3b of Li et al. \cite{Li2024}. Experimental PLQE, as reported in ref. \cite{Li2024}, is shown as symbols in Fig. \ref{exp_PLQY}a. Red symbols correspond to the control sample while blue symbols correspond to the dual sample. The performance improvement of the dual sample was attributed to accelerated radiative recombination facilitated by an increase in the exciton binding energy (from $3.9 \,meV$ for control sample to $13.9 \,meV$ for the dual sample \cite{Li2024}).\\

Using eq. \ref{k2}, the results for control sample in Fig. \ref{exp_fit} yields $k_1 = 2.1 \times 10^5 \,s^{-1}$ and $k_2^{'}=2.6 \times 10^{-11} \,cm^3s^{-1}$. The value for $k_1$ compares well with the estimate obtained using eq. \ref{k1} from the long tail of the transients ($k_1=1.96 \times 10^5 \,s^{-1}$). With the initial estimates for $k_1$ and $k_2$, a few other parameters could be back extracted from the PLQE measurements. For example, the experimental PLQE (see red symbols in Fig. \ref{exp_PLQY}) varies logarithmically with intensity. Using eq. \ref{eq:PLQY_u1}, we infer the presence of uniform trap density $N_T=9\times 10^{15} \,cm^{-3} eV^{-1}$ in the sample. Approximating eq. \ref{eq:PL_max1} as 
$PLQE_{max}=k_2^{'}/(k_2^{'}+2\sqrt{k_1k_3})$ and with the back extracted values for $k_1$, $k_2^{'}$, and $PLQE_{max}=70\%$ we find an initial estimate for $k_3\approx 10^{-28}\,cm^6s^{-1}$. As the $PLQE_{max}$ of this sample is relatively low, we may assume that the excitonic contribution (i.e., $k_x/n_{ep}$) to $k_2^{'}$ is not significant. Given $n_{ep}=9.6\times 10^{17} \,cm^{-3}$, we find that the $k_x \approx 10^6 \,s^{-1}$ satisfies this criterion.\\

The back extracted parameters for the control sample are provided in Table 1. With the reported exciton binding energy $E_b$, $n_{ep}$ was evaluated through Saha relation (eq. \ref{eq:saha}, Methods section). The dashed lines in Fig. \ref{exp_PLQY}a correspond to the PLQE as predicted by eq. \ref{eq:PL_max1}a - i.e., without the influence of traps. The solid lines correspond to numerical simulation results of eq. \ref{eq:traprate} under steady state conditions - i.e., with the influence of traps explicitly taken into account. For low intensities, the role of traps becomes significant and consequently the numerical solution of the general formalism (eq. \ref{eq:traprate}) anticipates the experimental results very well. The back extracted $k_2$ was assigned to both radiative and non-radiative components ($k_2=k_{2,rad}+k_{2,nonrad}$) to achieve a good comparison with the experimental PLQE (see Table 1, Methods section). Accordingly, the intensity dependent recombination components for the control sample is shown in Fig. \ref{exp_PLQY}b. We note the presence of significant non-radiative recombination and negligible contribution from excitons (i.e., with $E_b =3.9 \,meV)$ for this sample.\\ 

The parameters for the dual sample are obtained as follows: The parameters $k_{rad}$ and $k_3$ are kept the same as that of the control sample as there are no strong reasons to modify them. As its $PLQE_{max}$ is almost 100\%, we set $k_{2,nonrad}=0$ for the dual sample. The $PLQE_{max}$ of dual sample occurs at a lower intensity as compared to the control sample. This indicates that $k_1$ for the dual sample should be much smaller than that of control sample. Estimates for $k_1$ and $k_x$ were obtained using eq. \ref{eq:PL_max1}b. The predicted PLQE using eq. \ref{eq:PL_max1}a, shown as blue dashed line, compares well with the experimental results  - which indicates the presence of much lower trap density in the dual sample. The appropriate value for $N_T$ was obtained through numerical solution of eq. \ref{eq:traprate} and its comparison with experimental results.\\

Figure \ref{exp_PLQY}c shows the intensity dependence of the various recombination mechanisms for the dual sample. As evident from the back extracted parameters (see Table 1), we find that the excitonic component (i.e., with $E_b=13.9 \,meV$) to be comparable to the radiative component (i.e., $k_x/n_{ep} \approx k_{rad}$) which results in increased photo-emission. This increase in effective radiative recombination rate reduces the relative contribution of both non-radiative as well as Auger recombination components. The non-radiative component is itself significantly suppressed due to the much reduced $k_1$ - a reflection of the improved material quality. The same is also evident from the reduced bulk trap density associated with the dual sample (see Table 1). The back extracted parameters (see Table 1) compare well with those estimated through open circuit voltage transients of a solar cell limited by bimolecular recombination \cite{Hossain2024}.
\section{Discussions}
Several aspects are noteworthy in the results and insights shared in this manuscript. First, eqs. \ref{k1}-\ref{eq:PLQY_u1} constitute a coherent scheme to back extract the relevant parameters from PL measurements. This scheme do not rely on detailed curve-fitting of PL transients with eq. \ref{eq:sim_rate}. As a consequence, the back extracted parameters consistently interpret the device performance. Further, the proposed scheme allows physics aware design of experiments to probe the performance limiting factors. \\

As mentioned earlier, multiple reports in recent literature attributes the parameter $k_1$ in eq. \ref{eq:sim_rate} to monomolecular free carrier recombination and excitonic recombination \cite{herz_exciton,Li2024}. Indeed, ref. \cite{Li2024} justifies the improved performance of dual device (i.e., the PLQE improves from 70\% to 96\%) in part due to an increase in $k_1$ from $10^4 \,s^{-1}$ to $10^5 \,s^{-1}$. This approach is conceptually incorrect and is not helpful in elucidating the underlying device physics. Our approach clearly delineates and quantifies excitonic component from the monomolecular recombination. As per our analysis, the improvement in PLQE is, apart from the excitonic contribution, due to a decrease in $k_1$ from $2\times 10^5 \,s^{-1}$ to $3\times 10^4 \,s^{-1}$ and a reduction in the trap density - which is intuitive. In addition, the excitonic contribution is also clearly evident in terms of changes in the parameter $k_x$ and $E_b$ dependent $n_{ep}$. \\

Our analysis also leads to consistent estimates for the bimolecular recombination parameter. The parameter $k_{rad}$ for the both the samples is the same (see Table 1) and compares well with the value obtained through intensity dependent open circuit voltage transients \cite{Hossain2024} and fundamental radiative limits (for $E_g=1.6 eV$) \cite{bolink_eqeel,sumanshu_apl}. However, the effective bimolecular radiative recombination rate $k_2{'}=k_{rad}+k_x/n_{ep}$ is about $5\times 10^{-11} \,cm^3/s$ for the dual sample which along with much reduced $k_1$ results in very high PLQE. The $k_2$ value back extracted from PL transients, as reported in ref. \cite{Li2024}, is about $10^{-9} \,cm^3/s$. This clearly seems to be an artifact of the curve-fitting exercise and seems unrealistic (unless the $k_x$ value is experimentally shown to be of the order of $10^9 \,s^{-1}$ and/or a two orders of magnitude reduction in $n_{ep}$).\\

Eq. \ref{eq:PL_max1}b conveys important trends to achieve near ideal results: (a) $k_{2,nonrad}$ should be minimized, (b) larger $k_x$ and/or smaller $n_{ep}$, and (c), the material quality should be such that $k_1 << (k_{2}^{'})^2/4k_3$. While criteria (a) is self-evident, (b) indicates that larger $E_b$ is desirable (with excitonic peak significantly shifted from the band-band emission). With typical values of $k_{rad} \approx 10^{-11} \,cm^{3}/s$ and $k_3 \approx 10^{-28} \,cm^{6}/s$, we find that $k_1$ should be lower than $10^5 \,s^{-1}$. The back extracted $k_1$ for the dual sample meets this criteria (see Table 1). Further, the eq. \ref{eq:PL_max1}b  (and Fig. \ref{exp_fit}b) also indicates that the parameter set estimated through curve-fitting of PL transients might not be useful to interpret device performance.\\

Back extraction of physical parameters from experimental data can be regarded as an 'inverse' problem - akin to hearing the shape of a drum \cite{Kac}. As a consequence, multiple parameter sets could be possible solutions of varying accuracy. Indeed, as predicted by eq. \ref{eq:PL_max1}b and supported by numerical solutions, we find that the reduced parameter $k_x/n_{ep}$ influences the PLQE and not the individual value of $k_x$ or $n_{ep}$ (i.e., within an order of magnitude of variation). For a given $k_x/n_{ep}$, a few combinations of other parameters also yield similar results. However, all of them are comparable in magnitude (i.e., with in a factor of 2) and hence the set of parameters provided in Table 1 and used in Fig. \ref{exp_PLQY} consistently describes the carrier - exciton dynamics.\\

The proposed methodology (eqs. \ref{eq:eff_rate1} and \ref{eq:traprate}) assumes spatial homogenity for all parameters and processes. However, the $\tau^{-1}$ vs. $n(0)$ results for the dual sample (see blue symbols in Fig. \ref{exp_fit}) indicate that this assumption might need to be revisited. The slope of this data set is $9.8 \times 10^{-11} \,cm^3/s$.  Using the parameters listed in Table 1 (i.e., $E_b = 13.9 \,meV$, $n_{ep}= 6.5 \times 10^{17} \,cm^{-3}$, $k_2^{'}=4.8 \times 10^{-11} \,cm^3/s$, and $k_3=10^{-28} \,cm^6/s$), the slope obtained from the experimental data is comparable to both $k_3n_{ep}=6.54\times 10^{-11} \,cm^3/s$ (see eq. \ref{k3}) and $2k_2^{'}=9.6\times 10^{-11} \,cm^3/s$ (i.e., with excitonic component, see eq. \ref{k2}). Meanwhile, the intercept ($2.3 \times 10^6 \,s^{-1}$) is more closer to $k_2^{'}n_{ep}=3\times 10^7 \,s^{-1}$ (see eq. \ref{k3}) than to $2k_1=6 \times 10^4 \,s^{-1}$ (see eq. \ref{k2}). However, the carrier density involved ($n(0) \approx 10^{17} \,cm^{-3}$) is lower than $n_{ep}$ where eq. \ref{k3} is strictly applicable. As such, it is not directly evident which model should be used to interpret this experimental trend.\\

The above described apparent puzzle may be resolved, at least partially, as follows:  We may assume that the excitonic contribution is of significance in a volume fraction of $\alpha$. A first order analysis indicates that the relevant rate equation is $\partial n/\partial t \approx -(k_1(1-\alpha)+\alpha k_2^{'}n_{ep}/2)n-(k_2(1-\alpha)+\alpha k_3n_{ep}/2)n^2$. Following the analysis of eq. \ref{k2} or eq. \ref{k3}, with $\alpha=0.1$, we find that the expected slope in this case is $2k_2(1-\alpha)+\alpha k_3n_{ep} \approx 5\times 10^{-11} \,cm^3/s$ while the intercept is $2k_1(1-\alpha)+\alpha k_2^{'}n_{ep} \approx 3\times 10^6 \,s^{-1}$. Both these values compare well with experimental results for the dual sample (see Fig. \ref{exp_fit}) which indicate that spatial heterogenity indeed plays a significant role. In addition, the scaling trends of initial PL ($PL(t=0)$ with intensity  (i.e., $PL(t=0) \propto G^\beta$, with $\beta<2$) also indicates the possibility of spatial heterogenity. These aspects are in tune with recent literature \cite{xing2017,He_lowT_phase,Li2024} and might of future research interest. 
\section{Conclusions}
In this manuscript, we have proposed a consistent methodology to extract the parameters from photoluminescence without resorting to global curve-fitting.  To this end, we relied on analytical models and numerical simulations which explicitly accounts for the relevant physical mechanisms associated with free carrier, excitons, and traps.  The predicted PLQE closely matches the experimental results which validates our methodology. This allows quantitative interpretation and analysis of performance limiting factors in perovskite optoelectronic devices. Further, the developed methodology is of immense potential towards design of experiments to probe the carrier-exciton dynamics in novel optoelectronic devices and hence could be of broad interest.\\
\section{Acknowledgement}
To be updated
PRN ackowledges National Center for Photovoltaics Research and Education (NCPRE) IIT Bombay.\\
\section{Methods}
\textbf{Detailed model and numerical solution}:
Equation \ref{eq:traprate} describes the free electron - exciton dynamics in the presence of mid-gap recombination centers, carrier trapping, bimolecular recombination, and Auger recombination\cite{PierretADF}.  Similar approach, although not as extensive as ours, was reported in earlier publications \cite{stranks_pra,srh+model,Mariano2020,Sun2023}. However, all of them attempted to back extract the parameters through direct comparison of numerical solutions to PL transients - unlike the methodology proposed in this work which relies on asymptotic analytical solutions for appropriate regimes.

\begin{subequations}
\begin{align}
     \frac{\partial n}{\partial t} &= G -k_1n-k_2(np-n_i^2)-k_3(n+p)(np-n_i^2)\nonumber \\
                                   &-k_{nx}np+k_{xn}n_x -C_{nT}+E_{nT}\\ 
      \frac{\partial n_x}{\partial t} &= k_{nx}np-k_{xn}n_x-k_{x}n_x\\
     \frac{\partial n_T}{\partial t} &= C_{nT}-E_{nT}\\ 
      p &= n+n_T \\
      PL &=k_{rad}(np-n_i^2)+k_xn_x\\
      PLQE &= \frac{k_{rad}(np-n_i^2)+k_xn_x}{G}
    \end{align}
    \label{eq:traprate}
\end{subequations}

Here, eq. \ref{eq:traprate}a describes the time evolution of $n$. The parameter $G$ denotes the optical carrier generation rate while the parameters $k_1$, $k_2$ and $k_3$ represent monomolecular (i.e., due to mig-d-gap recombination centers), bimolecular and Auger recombination of free carriers, respectively ($n_i$ is the intrinsic carrier concentration). The parameter $k_{nx}$ denotes the rate of formation of excitons from free carriers, and $k_{xn}$ denotes the formation of free carriers due to exciton dissociation. The bimolecular recombination could have radiative and non-radiative components ($k_2=k_{2,rad}+k_{2,nonrad}$). \\

The parameter $G$ is a function of the illumination intensity ($I$) as $G=fI/(W\times hc/\lambda)$, $W=60 \,nm$ is the sample thickness, $\lambda=445 \,nm$ is the wavelength of the incident photons, $h$ is the Planck's constant, $c$ is the velocity of light. Here we assume that the entire incident photons are absorbed (i.e., $f=1$). The parameters $W$ and $\lambda$ are as reported in ref. \cite{Li2024}.\\

Eq. \ref{eq:traprate}b describes the dynamics of excitons with $k_x$  being the radiative decay rate of excitons. Under detailed balance between free carriers and  excitons (rather the Saha equilibrium), we have $k_{nx}n^2=k_{xn}n_x$. Accordingly, the parameters $k_{nx}$ and $k_{xn}$ are inter-dependent through the exciton binding energy ($E_b$) as 
\begin{subequations}
\begin{align}
     \frac{k_{xn}}{k_{nx}}&=n_{ep}\\
      n_{ep}&=(2\pi\mu kT/h^2)^{3/2}e^{-E_b/kT}   
\end{align}
    \label{eq:saha}
\end{subequations}
where $\mu$ is the reduced mass (assumed as $0.2m_0$, where $m_0$ is the rest mass of electrons), and $kT$ is thermal energy.  \\

Carrier trapping at single-level traps is accounted for in eq. \ref{eq:traprate}a with terms involving trap activity like electron capture by unfilled traps ($C_{nT}=c_n(N_T-n_T)n$) and electron emission from filled traps  ($E_{nT}=c_nn_Tn_1$), where $N_T$ denotes the density of total traps. Eq. \ref{eq:traprate}c defines the rate of change of $n_T$, the density of filled traps. Here $c_n$ denotes the capture coefficient of electrons by filled traps. The parameter $n_1$ depend on the trap energy as $n_1=n_ie^{E_T/kT}$ where $E_T$ is the effective trap energy level. Different trap distributions like single-level, uniform, or exponential band tail states can be accounted for in this formalism. The monomolecular recombination ($k_1n$) could also be represented in terms of trap activity (i.e., with additional terms for hole capture and hole emission).\\

It is important to note that a separate rate equation for hole density is not needed. On the contrary, the influence of holes is accounted explicitly through the principle of charge neutrality - i.e., eq. \ref{eq:traprate}d indicates that the electrons could be either in the conduction band or in the filled traps, while the holes are in the valence band. The same equation could be appropriately modified to account for the presence of unintentional doping, if any. \\ 

 The photon emission due to free carriers and excitons is given by eq. \ref{eq:traprate}e.  If the exciton binding energies are low, the emission spectra due to free carrier recombination might have significant overlap with that of excitons. Indeed, ref. 1 indicates that an increase in exciton binding energy by  $10 \,meV$, results in  a red-shift of the peak emission by $4 \,nm$. Accordingly, the PL signal and PL quantum efficiency are given by eqs. \ref{eq:traprate}e and \ref{eq:traprate}f, respectively. These equations could be appropriately modified to consider the wavelength dependent emission.\\

Numerical solution of eq. \ref{eq:traprate} is achieved through fully implicit Backward Euler method. The results shared in Fig. \ref{exp_PLQY} were obtained through the numerical solution of eq. \ref{eq:traprate} (and not eq. \ref{eq:eff_rate1}) under steady state conditions. An important aspect in this regard is that eq. \ref{eq:traprate} do not assume Saha equilibrium between carriers and excitions while eq. \ref{eq:eff_rate1} indeed relies on the same. Table 1 lists the relevant parameters involved. The general scheme, under certain assumptions like  (a) quasi-steady state between excitons and free carriers, (b) quasi-steady state conditions for traps, (c) mid-level traps, and (d) $n=p$ reduces to eq. \ref{eq:eff_rate1}. Accordingly, we note that the $k_3$ parameter mentioned in these two formalisms should differ by a factor of 2, while ensuring that the Auger recombination component remains the same. \\

\begin{table}[h]
    \centering
    \begin{tabular}{|c|c|c|c|}
    \hline
        Parameter & Units & Control & Dual  \\
        \hline
         $k_1$ & $s^{-1}$ & $2\times 10^5$ & $0.3\times 10^5$\\
         \hline
         $k_{2,rad}$ & $cm^3s^{-1}$ & $2.4 \times 10^{-11}$ & $2.4 \times 10^{-11}$ \\
         \hline
         $k_{2,nonrad}$ &  $cm^3s^{-1}$ & $0.2 \times 10^{-11}$ & 0 \\
         \hline
         $k_{2}^{'}$ &  $cm^3s^{-1}$ & $2.6 \times 10^{-11}$ & $4.8 \times 10^{-11}$ \\
         \hline       
         $k_3$ & $cm^6s^{-1}$ &$ 10^{-28}$ &$10^{-28}$\\
         \hline
         $k_x$ & $s^{-1}$ & $10^6$& $1.6 \times 10^7$\\
         \hline
         $E_b$ & $meV$& 3.9 & 13.9\\
         \hline
         $n_{ep}$ & $cm^{-3}$ & $9.6 \times 10^{17}$ & $6.5 \times 10^{17}$\\
         \hline
         $k_{nx}$ & $cm^{3} s^{-1}$& $10^{-6}$ & $10^{-6}$\\
         \hline
         $k_{xn}$ & $s^{-1}$ & $k_{nx} n_{ep} $ & $ k_{nx} n_{ep}$\\
         \hline
         $N_T$ & $cm^{-3} eV^{-1}$& $9 \times 10^{15}$ & $1.25 \times 10^{15}$\\
         \hline
    \end{tabular}
    \caption{Parameters obtained through the proposed methodology (eqs. \ref{k1}-\ref{eq:PLQY_u1})and used in comparison between theory and experiments in Fig. \ref{exp_PLQY}}. 
    \label{tab:my_label}
\end{table}

\bibliography{apssamp}

\end{document}